\documentclass[aps,pra,preprint,groupedaddress]{revtex4-1}

\usepackage{graphicx}
\usepackage{bm}

\newcommand{\Rmnum}[1]{\expandafter\@slowromancap\romannumeral #1@}
\newcommand{\sech}{\mathrm{sech}}

\newcommand{\ket}[1]{|#1\rangle}

\newcommand{\X}{\mathrm{X}}
\newcommand{\XX}{\mathrm{XX}}

\newcommand{\Wcm}{\mathrm{W}/\mathrm{cm^2}}
\newcommand{\tnl}{\mathrm{tnl}}
\newcommand{\rad}{\mathrm{r}}
\newcommand{\thm}{\mathrm{thm}}
\newcommand{\rme}{\mathrm{e}}
\newcommand{\rmd}{\mathrm{d}}

\begin{document}

\title{Carrier dynamics in site- and structure-controlled InGaN/GaN quantum dots}

\author{Lei Zhang$^1$, Tyler A. Hill$^1$, Chu-Hsiang Teng$^2$, Brandon Demory$^2$, Pei-Cheng Ku$^2$, and Hui Deng$^1$}

\address{$^1$ Department of Physics, University of Michigan, 450 Church Street, Ann Arbor, MI 48109, USA}
\address{$^2$ Department of Electrical Engineering and Computer Science, University of Michigan, 1301 Beal Avenue, Ann Arbor, MI 48109, USA}

\date{\today}

\begin{abstract}
We report on the carrier dynamics in InGaN/GaN dot-in-nanowire quantum dots revealed by systematic mapping between optical properties and structural parameters of the quantum dots. Such a study is made possible by using quantum dots with precisely controlled locations and sizes. We show that the carrier dynamics is governed by two competing mechanisms: 1) excitons are protected from surface recombination by a potential barrier formed due to strain-relaxation at the sidewall surface; 2) excitons can overcome the potential barrier by tunnelling and thermal activation. This carrier dynamics model successfully explains the following surprising experimental findings on individual quantum dots. Firstly, there exist strong statistical correlations among multiple optical properties of many individual quantum dots, despite variations of these properties resulting from inevitable structural variations among the quantum dots. Secondly, the antibunching property of quantum dot emission exhibits abnormal ladle-shaped dependence on the decay time and temperature. Our results can guide the way toward nitride-based high temperature single-photon emitters and nano-photonic devices.
\end{abstract}


\maketitle

\section{Introduction}
\label{sec:intro}

Nitride-based quantum dots (QDs) are one of the most promising candidates for achieving on-chip electrically-driven single-photon sources at room temperature \cite{Kako2006b, Deshpande2013a, Holmes2014}. 
The optical performance of these devices, such as their quantum efficiency (QE), operation speed and single-photon purity, critically depends on the carrier dynamics, such as the radiative and nonradiative decay rates of the carries. The carrier dynamics, in turn, is determined by nano-scale structural parameters, such as the diameter, thickness, material compositions and surface properties. 
Therefore, the key to improve the performance is to know what the structural parameters are and how they impact the performance. 
However, it is nearly impossible to directly measure all the relevant structural parameters of every single QD non-destructively with sufficient accuracy using current technologies \cite{Rigutti2014}.
Furthermore, the correlation between the optical properties and the structural parameters was obscure in previous studies in which the QDs were typically formed at random positions with large structural inhomogeneity \cite{Edwards2004, Holmes2011a}.

Alternatively, we have recently demonstrated high-quality site-controlled InGaN QDs whose structural parameters were controlled to the limit of the state-of-the-art growth and nano-fabrication technologies \cite{Lee2011, Zhang2013b}.
Using these QDs, we are able to systematically study the optical properties of QDs with diameters ranging from 19~nm to 33~nm, using time-integrated and time-resolved photoluminescence (PL) spectroscopy as well as second-order correlation ($g^{(2)}$) measurements at temperatures from 10~K to 120~K. 
Such a study allows us to observe strong correlations between the structural parameters and optical properties and, based on which, extract the underlying carrier dynamics in these nitride-based QDs.

In addition, we show that opposed to the common practice assuming 100\% quantum efficiency at low-temperatures, quantum-tunnelling leads to significant surface recombination in some QDs even at temperatures down to 10~K. Furthermore, contradictory to conventional expectations, we show that for our QDs the best photon-antibunching does not always occur in the brightest QDs or at the lowest temperatures. The above observations are intuitively explained by our carrier dynamics model, providing guidance for future improvement of single-photon sources based on III-nitride QDs.

This article is organized as following.
Section~\ref{sec:experiment} presents the QD sample stucture and the optical setup used to study it.
Section~\ref{sec:carrier_dynamics_principles} presents the principles of the carrier dynamics and explains its strong dependence on structural parameters and, therefore, the necessity of using site- and structure-controlled QDs to study it. In Secs.~\ref{sec:potential} and \ref{sec:decay_rates} we establish the main control mechanisms and parameters of carrier dynamics using QD ensembles of different diameters.
Section~\ref{sec:potential} identifies a lateral confinement potential profile for the excitons and extracts its analytical form based on the diameter dependence of the PL energy of the QDs.
Section~\ref{sec:decay_rates} extracts the key parameters needed to model the radiative, tunnelling and thermal decay rates of excitons in our QDs, based on the diameter and temperature dependences of the  PL decay time of the QDs.
In Secs.~\ref{sec:statistics} and \ref{sec:g2} we extend the carrier dynamics model to describe properties of individual QDs.
Section~\ref{sec:statistics} identifies the sources of inhomogeneities in PL properties among individual QDs of closely matched structural parameters, and explains the statistical correlations among these PL properties using the carrier dynamics model.
Section~\ref{sec:g2} includes the biexciton dynamics into the model by considering exciton-exciton interactions and explains the peculiar decay time and temperature dependence of the degree-of-photon-antibunching.
Finally, Sec.~\ref{sec:conclusion} summarizes the main findings.

\section{Sample structure and optical setup}
\label{sec:experiment}

\begin{figure}
\includegraphics[width=0.9\textwidth]{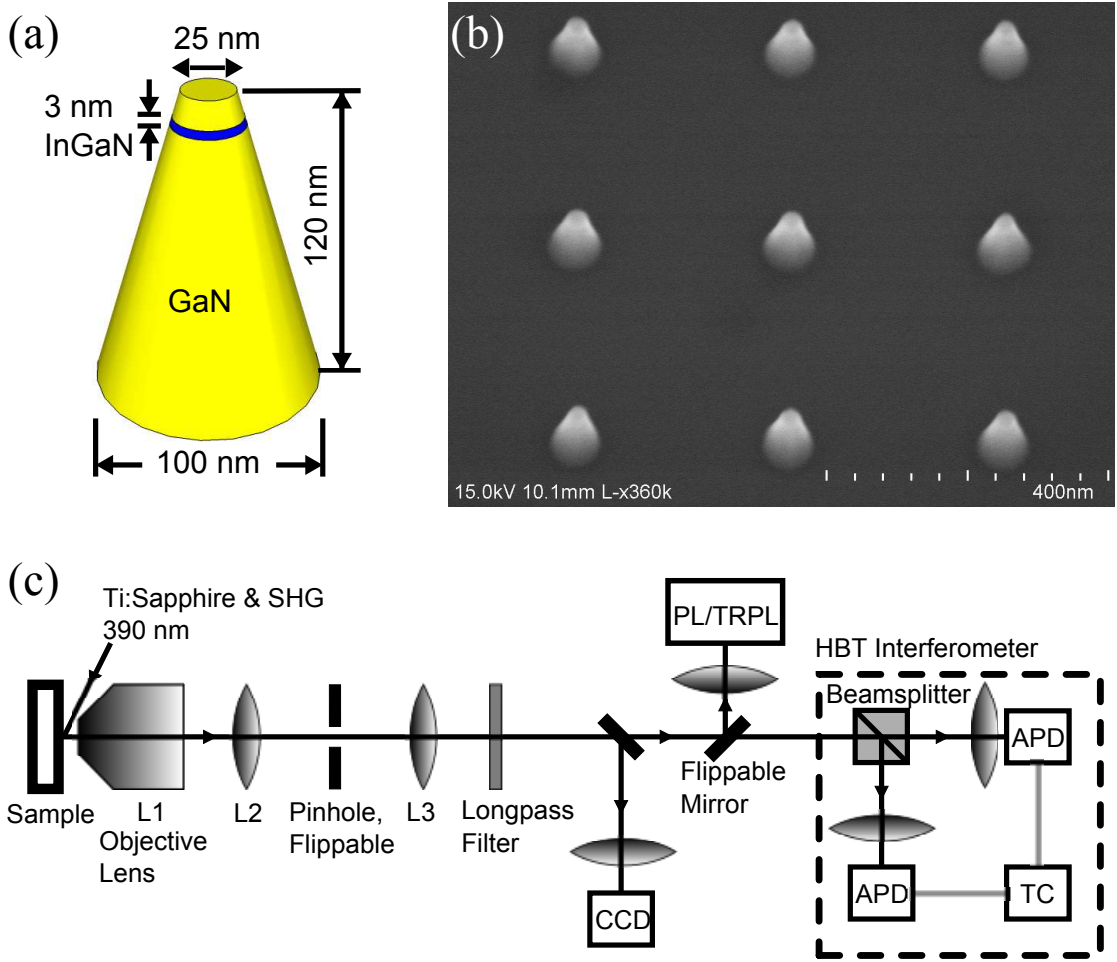}
	\caption{
	(a) The schematic plot of a single GaN nanopillar containing an InGaN nanodisk of diameter $D=25$~nm.
	(b) The SEM image of a part of a dense arrays of QDs of diameter $D = 25$~nm and dot-to-dot separation 300~nm. The viewing angle is 45$^\circ$.
	(c) The schematic plot of the optical setup used in this work, in which SHG stands for second harmonic generation; CCD, charge-coupled device; PL/TRPL, photoluminescence and time-resolved photoluminescence measurement setup; APD, avalanche photo-diode; and TC, time correlator.}
	\label{fig:sample_and_optics}
\end{figure}

The QDs studied in this work were fabricated via electron-beam patterning and plasma etching a planar single InGaN/GaN quantum well (QW) \cite{Lee2011a}. Details of the fabrication processes can be found in \cite{Lee2011a, Zhang2013b}. Each individual QD was made of a 3~nm thick In$_{0.15}$Ga$_{0.85}$N nanodisk in a 120~nm tall GaN nanopillar, as illustrated in Fig.~\ref{fig:sample_and_optics}(a). The nanodisk had a 10~nm thick GaN at the top and had sidewalls exposed in air. We studied multiple dense and sparse arrays of QDs with InGaN nanodisk diameters ranging from 19 to 32~nm. Each dense array consisted of $100 \times 100$ QDs with a 300 nm inter-dot separation (Fig.~\ref{fig:sample_and_optics}(b)) for QD ensemble study. Each sparse array (not shown) consisted of $10\times 10$ QDs with a 5~$\mu$m inter-dot separation for individual QD study. The diameters of all InGaN QDs in the same array were nominally the same with $\sim 2$~nm standard deviation. 

The optical setup used in this work is illustrated in Fig.~\ref{fig:sample_and_optics}(c). The sample was placed in a temperature stabilized He-flow cryostat. It was excited from an angle 55$^\circ$ from the normal direction with a 390~nm laser obtained by frequency doubling a 780~nm Ti:Sapphire laser with a 150~fs pulse duration and an 80~MHz repetition rate. We studied the PL and time-resolved PL (TRPL) properties of QDs using a confocal microscope setup composed of an objective lens (L1, $f=5$~mm, 0.6 NA) and a  pair of confocal lenses (L2 and L3, $f=75$~mm). The setup was switchable from ensemble to single QD measurements by  placing a 25~$\mu$m pinhole at the confocal plane. The pinhole was used to spatially selected the luminescence from a $<2~\mu$m region on the sample, hence, collect the emission from only one QD from the sparse array with a 5 $\mu$m inter-dot separation. A longpass filter was used to remove the scattered laser and only let the PL from QDs through. A CCD camera was used to monitor the position of QDs through their PL. The PL spectrum was measured using a spectrometer (part of the PL/TRPL box in Fig.~\ref{fig:sample_and_optics}(c)) with a 0.6~meV (0.08~nm) spectral resolution at $\sim 3$~eV (400~nm). The second-order correlation ($g^{(2)}$) function of single QDs was measured using a Hanbury Brown-Twiss (HBT) interferometer \cite{Brown1956} composed of a 50:50 beamsplitter, two avalanche photo-detectors (APDs) and a time-correlator (TC). One of the APD-TC arms was also used for the TRPL measurement (not shown in Fig.~\ref{fig:sample_and_optics}) with a synchronization signal from the Ti:Sapphire laser. The APD-TC system's instrument response function (IRF) at 400~nm had a $\sim 0.2$~ns full-width-at-half-maximum (FWHM), which corresponds to the time resolution of the TRPL and $g^{(2)}$ measurements.

\section{Principles of carrier dynamics}
\label{sec:carrier_dynamics_principles}

In this section, we summarize the principles of the carrier dynamics and how they were identified by using our site- and structure-controlled QDs. The detailed deduction of the model based on the experimental data will be presented in later sections. In our QDs the nonradiative decay is determined by two competing processes: on one hand, the strain-relaxation at the sidewall forms a confinement potential protecting excitons from recombining with the nonradiative surface states; on the other hand, excitons can overcome the potential barrier through tunnelling and thermal activation. 

\begin{figure}
\includegraphics[width=0.9\textwidth]{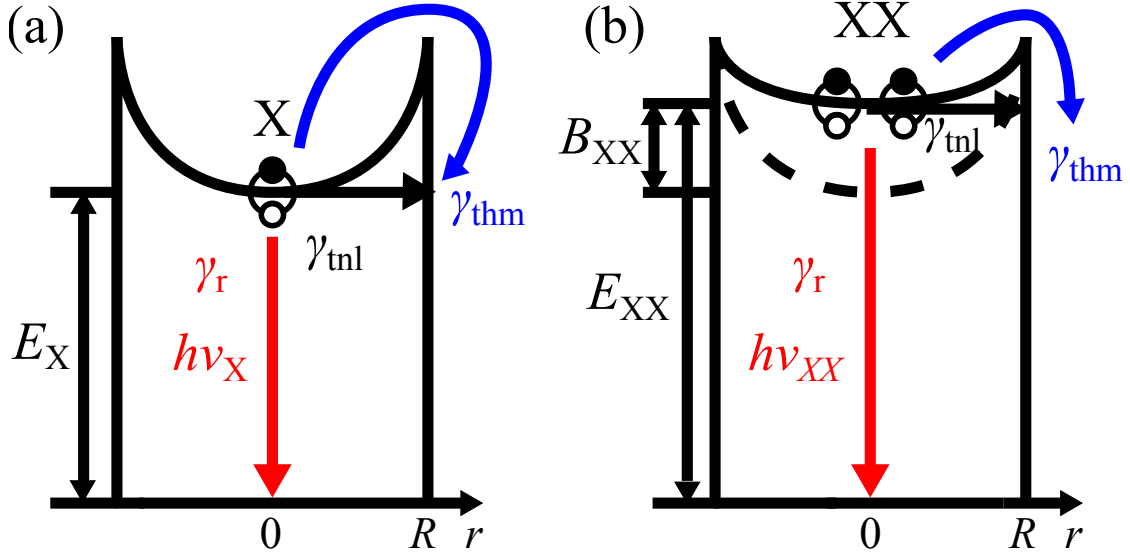}
	\caption{
	(a) Exciton decay model. A lateral potential profile is formed mainly due to the reduction in the piezoelectric field at the nanodisk sidewall. An exciton decays mainly through three channels: the radiative decay and the tunnelling and thermal surface nonradiative decay, whose decay rates are denoted as $\gamma_\rad$, $\gamma_\tnl$ and $\gamma_\thm$, respectively. The latter two processes need to overcome the lateral potential barrier.
	(b) Biexciton decay model. The presence of another exciton lowers the potential barrier experienced by any of the two excitons due to the screening effect, leading to a lower biexciton QE compared to exciton QE.}
	\label{fig:model}
\end{figure}

In an etched InGaN/GaN nanodisk, the lateral confinement potential for the exciton is formed due to strain-relaxation at the sidewall, as illustrated in Fig.~\ref{fig:model}(a). This is manifested as a continuous blueshift of QD PL energy with the reduction of the diameter (Fig.~\ref{fig:potential}), which shows the piezoelectric field is the determining factor of the potential profile, as explained in \cite{STRAIN}. The larger strain-relaxation at the sidewall compared to the center of the nanodisk leads to a weaker piezoelectric field and, thus, higher exciton energy. The resulting confinement potential protects excitons from reaching the detrimental sidewall surface, and thus plays a critical role in exciton dynamics. The potential profile will be obtained in Sec.~\ref{sec:potential}.

The exciton dynamics in a QD is governed by its radiative and nonradiative mechanisms. We will examine them in Sec.~\ref{sec:decay_rates} by the diameter and temperature dependence of the PL intensity and decay time of QD ensembles. Here we present main principles of radiative and nonradiative decay below:

The radiative decay rate $\gamma_\rad$ of an exciton in an InGaN/GaN nanodisk is determined by its oscillator strength, the local density of photon states (LDPS) and the temperature. The oscillator strength increases due to strain relaxation, whereas the LDPS decreases as the nanodisk diameter is reduced \cite{STRAIN}. Meanwhile, since an exciton can be thermally scattered out of the radiation zone in the momentum space, the averaged $\gamma_\rad$ is expected to decrease as temperature increases \cite{Feldmann1987, Andreani1991}. The relative $\gamma_\rad$ is reflected by the ratio of the PL intensity and the decay time, as will be explained in Sec.~\ref{ssec:gamma_r}.

The nonradiative decay of an exciton can occur at the surface and in the volume of a QD. The volume recombination includes the Shockley-Read-Hall \cite{Shockley1952} and Auger \cite{Klimov2000a} processes. In our QDs, the volume recombination was negligible, because at any given temperature the planar QW on the same sample had a measured total decay time over ten times longer than that observed from the QDs investigated in this work.

The surface nonradiative recombination is of major concern for nanosturctures with large surface-to-volume ratios. There are two types of surface nonradiative decay processes based on the way excitons overcome the potential barrier and reach the sidewall surface: the tunneling decay and thermal decay, as illustrated in Fig.~\ref{fig:model}(a), which is analogous to the field emission and thermionic emission in the standard Schottky barrier theory \cite{sze2006physics, Padovani1966}. 

The tunnelling decay is the dominant nonradiative decay mechanism at low temperatures in our QDs. It is due to the tunnelling of the exciton through the potential barrier to the surface, and is determined by the wavefunction overlap of the exciton and surface states. Hence, the tunnelling decay rate $\gamma_\tnl$ is finite even at zero temperature as shown in Sec.~\ref{ssec:gamma_tnl}. It strongly depends on the potential barrier profile and only has a weak temperature dependence, similar to the thermionic-field-emission component in the Schottky barrier theory \cite{sze2006physics}. 

The thermal decay becomes the dominant nonradiative decay mechanism as the temperature increases. It is due to the thermal activation of the exciton over the potential barrier to the surface. Therefore, the thermal decay rate $\gamma_\thm$ depends strongly on the temperature as well as the potential barrier profile as shown in Sec.~\ref{ssec:gamma_thm}.

Experimentally characterizing the radiative, tunnelling and thermal decay independently is, however, far from straightforward, since only the total decay rate and PL intensity can be directly measured. 
Each of these decay rates varies differently with the potential profile, while the potential profile is determined by multiple structural parameters of the QD, such as the diameter, thickness and indium mole fraction. 
Therefore, to obtain the dependence of each decay mechanism on each structural parameter, one would have to measure the change of the optical properties while varying each structural parameter separately.
This cannot be done with QDs self-assembled at random sites, whose structural parameters are often correlated with each other and suffer from large inhomogeneity.

On the other hand, using the site- and structure-controlled QDs described in Sec.~\ref{sec:experiment}, we can measure the dependence of the optical properties on one of the structure parameters while keeping the fluctuations in others minimal. This enables us to extract the decay rates and their dependence on various structural parameters, as we discuss in Sec.~\ref{sec:decay_rates}. 

Despite the control of structural parameters, small fluctuations in these parameters are inevitable. This leads to inhomogeneities in the optical properties of QDs with the same nominal structural parameters, as shown in Sec.~\ref{ssec:inhomogeneity}. However, with the tight control of the structural parameters, it turns out that the inhomogeneities in all measured optical properties are strongly correlated with each other, as shown in Sec.~\ref{ssec:correlation}. As a result, variations in different optical properties can all be modeled by the variation of one parameter, which is identified as the sidewall potential barrier height.

The above carrier dynamics model describes well the intensity and decay time of the QD emission, which is dominated by the single-exciton emission at the low excitation intensities used in this work. However, to describe the photon-antibunching property, which is determined by both the exciton and biexciton dynamics \cite{Nair2011a}, biexciton dynamics needs to be included.

The biexciton emission is different from exciton emission by its binding energy, which is typically negative in InGaN QDs due to the repulsive exciton-exciton Coulomb interaction between the two excitons \cite{Schomig2004, Martin2005, Simeonov2008, Winkelnkemper2008, Bardoux2009, Sebald2011, Amloy2012, Chen2012, STRAIN}. This effectively leads to a lowering of the potential barrier for a biexciton by its binding energy, which can be extracted from the overall spectral linewidth, as explained in Sec.~\ref{ssec:bxx}. This allows us to explain the peculiar correlation between the photon-antibunching and other PL properties in Sec.~\ref{ssec:g2_vs_tau} and the abnormal temperature dependence of the photon-antibunching in Sec.~\ref{ssec:g2_vs_T}.

\section{The lateral potential barrier profile}
\label{sec:potential}

As explained in the preceding section, the carrier dynamics in an InGaN/GaN nanodisk strongly depends on the exciton potential profile in the nanodisk. Therefore, the first step towards modeling the carrier dynamics is to identify the sources of the potential profile. 


There are several mechanisms that could lead to an exciton potential profile in the InGaN/GaN nanodisk: the non-uniform distributions of the deformation potential as well as the spontaneous and piezoelectric polarizations due to the uneven strain distribution from the center to the sidewall of the nanodisk \cite{Bocklin2010a}. Note that we do not consider the contribution of the surface Fermi-level pinning \cite{Veal2006, Holmes2011a}, because it bends the conduction and valance bands in the same way and, hence, has negligible effects on the exciton potential profile.

\begin{figure}
\includegraphics[width=0.9\textwidth]{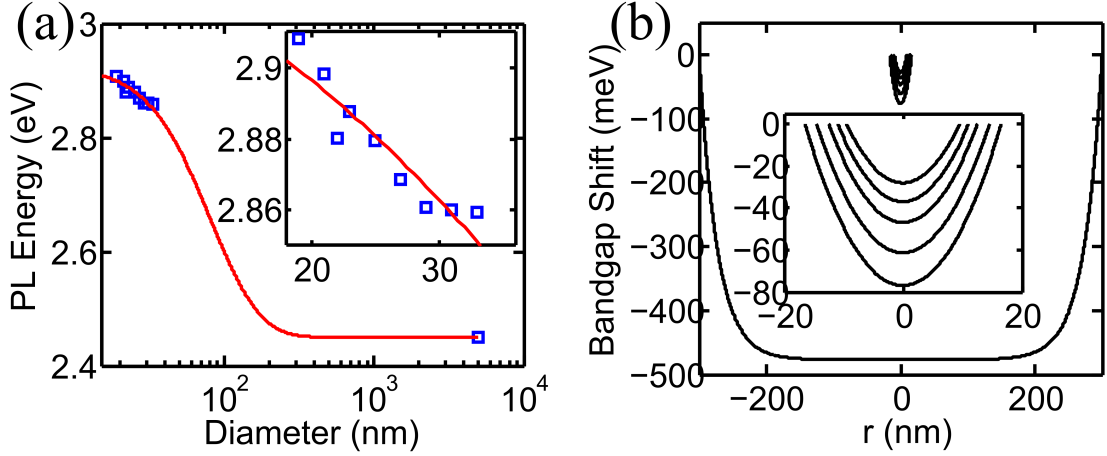}
	\caption{
	(a) PL energy $E$ vs. QD diameter $D$ of nine dense arrays of QDs with diameters ranging from 19~nm to 33~nm and a 5~$\mu$m diameter QW taken at excitation intensity $P = 1~\Wcm$. The inset is a zoomed-in version of the nine dense QD arrays. The solid line is a fitting curve based on Eq.~(\ref{equ:E_vs_D}).
	(b) The exciton potential profiles of QDs of diameter $D = 19$, 22, 25, 29, 33 and 600~nm calculated using Eq.~(\ref{equ:phi_prime}). The inset is a zoomed-in version of the five smaller diameters.}
	\label{fig:potential}
\end{figure}

To identify the source of the exciton potential profile we follow the same procedure as in \cite{STRAIN}. We measure the PL energy of nine dense arrays of QD-nanodisks with diameters varying from 19~nm to 33~nm as well as a QW-nanodisk with 5~$\mu$m diameter on the same sample at a temperature of 10~K. We use a very low excitation intensity $P = 1~\Wcm$ to avoid significant screening of the electric field. As shown in Fig.~\ref{fig:potential}(a), the PL energy blueshifts as the diameter of the QD reduces from 5~$\mu$m to 19~nm, consistent with previous studies \cite{Kawakami2006, Holmes2009, Hsueh2005, Chen2006a, Kawakami2010, Ramesh2010, STRAIN}. 

The large amount of blueshift ($\sim 500$~meV) from QW to QD shows that the piezoelectric polarization gives the dominant contribution to the exciton potential profile in our QDs \cite{STRAIN}. The piezoelectric polarization lowers the bandgap due to the quantum-confined Stark effect \cite{Takeuchi1997}. Therefore, reducing the diameter decreases the overall strain and piezoelectric polarization, leading to an increase in the bandgap and, correspondingly, the exciton energy. 
In contrast, the deformation potential increases the bandgap \cite{Bocklin2010a} and would lead to a redshift of the PL energy with decreasing the nanodisk diameter. The spontaneous polarization also cannot explain the blueshift, since it is largely independent of the strain and, therefore, the nanodisk diameter. 

The predominance of the piezoelectric polarization in shaping the potential profile suggests a lower exciton potential at the center of the nanodisk, as illustrated in Fig.~\ref{fig:model}(a). This is because the higher residue strain at the center of the nanodisk leads to stronger piezoelectric field and Stark redshift compared to at the sidewall. Following a phenomenological model in \cite{STRAIN}, the resulting potential profile at radial position $r$ can be described as:
\begin{equation}
\phi'(r) = - B_m [1-\sech (\kappa D/2) \cosh (\kappa r)]. 
\label{equ:phi_prime}
\end{equation}
Here, $B_m$ represents the amount of exciton energy shift from an unstrained QW to an fully-strained QW, $1/\kappa$ is the characteristic length of the region from the InGaN nanodisk sidewall where the compressive strain of the InGaN layer is relaxed, and $D$ is the nanodisk lateral diameter. Due to the predominance of the piezoelectric polarization over the deformation potential, $B_m$ is always positive, i.e. $\phi'(0)<\phi'(D/2)$.

At low excitation intensities, the diameter-dependent PL energy $E(D)$, corresponding to the exciton energy at $r=0$, can be written as:
\begin{equation}
E = E_0 - B_m[1-\sech (\kappa D/2)].
\label{equ:E_vs_D}
\end{equation}
Here, $E_0$ is the exciton energy at $r=D/2$ where the strain is considered as being fully relaxed. Therefore, it represents the bandgap of an unstrained QW or a nanodisk of $D \rightarrow 0$.

The characteristic strain relaxation parameters $B_m$ and $\kappa$ as well as $E_0$ are obtained by using Eq.~(\ref{equ:E_vs_D}) to fit the measured $E(D)$ data, as shown in Fig.~\ref{fig:potential}(a). We obtain, for this particular sample, $E_0= 2.93$~eV, $B_m = 477$~meV, and $1/\kappa = 27$~nm. With $B_m$ and $\kappa$, we use Eq.~(\ref{equ:phi_prime}) to calculate the exciton potential profile for QDs with different diameters as shown in Fig.~\ref{fig:potential}(b).

For describing the carrier dynamics, only the shape of the potential profile matters. For convenience in later discussions, we shift the $\phi'(r)$ in Eq.~(\ref{equ:phi_prime}) by an $r$-independent constant to obtain $\phi(r)$ so that $\phi(r=0) = 0$:
\begin{equation}
\phi(r) = B_m \sech (\kappa D/2)[\cosh (\kappa r)-1]. 
\label{equ:phi}
\end{equation} 
For a QD of diameter $D$, $\phi(r)$ reaches the maximum and minimum at $r=D/2$ and $r=0$, respectively, i.e. the potential barrier height is: \begin{equation}
\phi_B = \phi(D/2) - \phi(0) = B_m[1-\sech(\kappa D/2)].
\label{equ:phi_B}
\end{equation}

\section{Decay rates}
\label{sec:decay_rates}

As explained in Sec.~\ref{sec:carrier_dynamics_principles}, an exciton in an QD mainly undergoes three decay processes: the radiative, tunnelling and thermal decay, whose rates are denoted as $\gamma_\rad$, $\gamma_\tnl$ and $\gamma_\thm$, respectively (Fig.~\ref{fig:model}(a)). In this section, we show that to consistently explain all measured diameter and temperature dependence of the total decay time and PL intensity in this study, the total decay rate $\gamma = \gamma_\rad+\gamma_\tnl + \gamma_\thm$ has to be expressed as:
\begin{equation}
\gamma = \gamma_{\rad 0}( 1 - \rme^{- \Delta/k_BT}) 
+ \frac{c_1}{D} \rme^{-\frac{2\sqrt{2m}}{\hbar}\int \limits_0^{D/2} \sqrt{\phi(r)} \rmd r} 
+ \frac{c_2 v_B}{D} \rme^{-c_3\phi_B/k_BT}.
\label{equ:gamma}
\end{equation}
The three terms on the right-hand-side correspond to $\gamma_\rad$, $\gamma_\tnl$ and $\gamma_\thm$, respectively.

We show in Sec.~\ref{ssec:gamma_r}, in the $\gamma_\rad$ term, $\gamma_{\rad 0}$ is the radiative decay rate at temperature $T \rightarrow 0$, it is independent of the QD diameter $D$ for 21~nm $<D< 33$~nm; $k_B$ is the Boltzmann constant; and $\Delta$ is the exciton homogeneous linewidth limited mainly by impurity and interface scattering.  $\gamma_{\rad}$ decreases as $T$ increases, due to the thermal scattering of excitons out of the radiation zone \cite{Feldmann1987}.
We show in Sec.~\ref{ssec:gamma_tnl}, in the $\gamma_\tnl$ term,  $c_1$ is a unit conversion factor reflecting the probability of the surface recombination; $m$ is the exciton effective mass; and $\hbar$ is the reduced Planck constant. The exponential factor $\rme^{-\frac{2\sqrt{2m}}{\hbar}\int \limits_0^{D/2} \sqrt{\phi(r)} \rmd r}$ is proportional to the probability for a ground state exciton to tunnel through the potential profile $\phi(r)$ to the sidewall surface; the $1/D$ factor takes into account the surface-to-volume ratio.
We show in Sec.~\ref{ssec:gamma_thm}, in the $\gamma_\thm$ term, $c_2$ is a scaling factor similar to $c_1$ but with a different unit; $c_3$ is a factor that adjust the potential barrier height to account for the temperature-dependent part of the tunneling decay; the exponential factor $\rme^{-c_3\phi_B/k_BT}$ is the thermal population of excitons with kinetic energy higher than $c_3 \phi_B$; $v_B$ is defined as $v_B=\sqrt{2(c_3\phi_B + k_B T)/m}$, the average velocity of excitons with kinetic energies higher than $c_3 \phi_B$. 

\begin{figure}	
\includegraphics[width=0.9\textwidth]{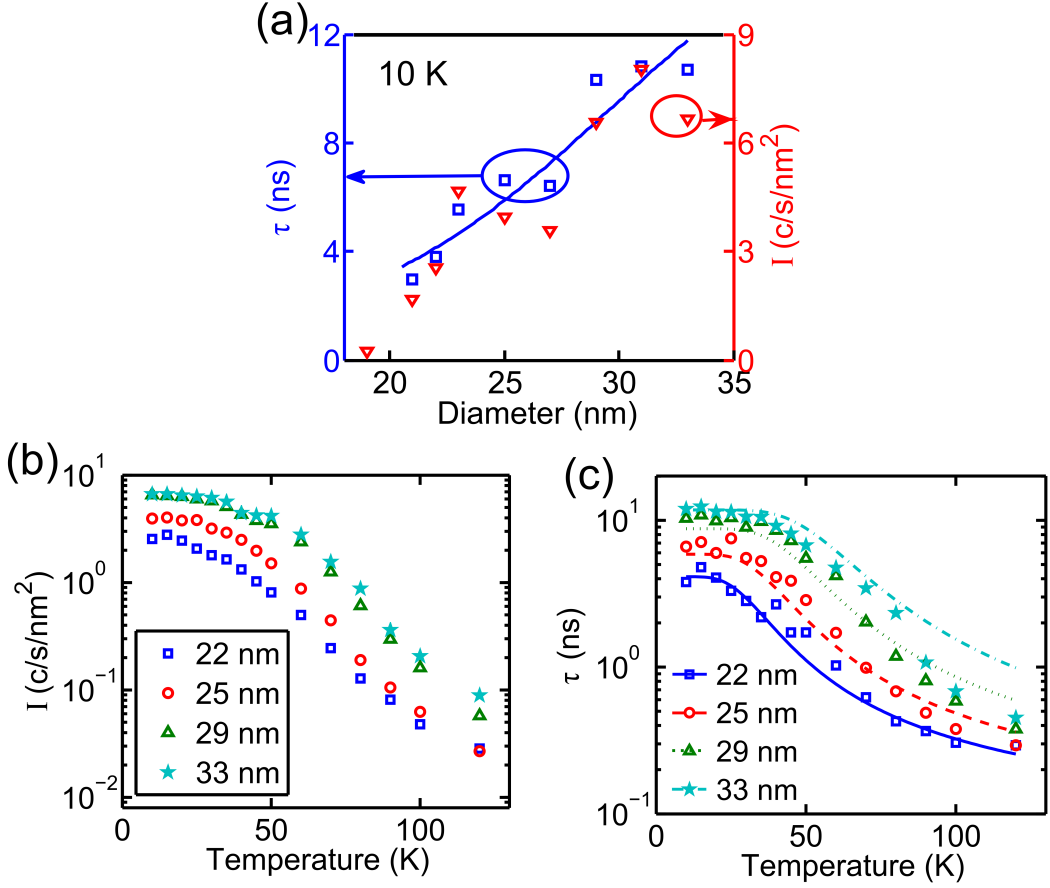}
	\caption{
	(a) The total decay time $\tau$ (square, left axis) and PL intensity per unit InGaN area $I$ (triangle, right axis) vs. QD diameter $D$ dense arrays at 10~K. The $\tau(D)$ data is fitted using Eq.~(\ref{equ:gamma_T0}), yielding $\gamma_{\rad 0} = 59$~MHz and $c_1 = 15$~m/s.
	(b) The $I(T)$ data of four dense arrays with QD diameter $D = 22$~nm (blue square), 25~nm (red circle), 29~nm (green triangle), 33~nm (cyan star) nm, respectively.
	(c) The $\tau(T)$ data of the same four arrays described in (b), which are simultaneously fitted using Eq.~(\ref{equ:gamma}), yielding $c_2 = 2 \times 10^{-3}$ and $c_3 = 0.33$, as shown by blue solid, red dash, green dot and cyan dash-dot curves. The fitting took into account the instrument time-resolution of 0.2~ns (Sec.~\ref{sec:experiment}).
	}
	\label{fig:decay_rates}
\end{figure}

The parameters $\phi(r)$ and $\phi_B$ in Eq.~(\ref{equ:gamma}) have been obtained in Sec.~\ref{sec:potential}. The remaining five unknown parameters in Eq.~(\ref{equ:gamma}), $\gamma_{\rad 0}$, $\Delta$, $c_1$, $c_2$ and $c_3$, can be obtained from carefully designed control experiments. These include measuring the diameter and temperature dependence of the PL intensity $I$ and the total decay time $\tau = 1/\gamma$ of QD arrays with tightly controlled structural parameters, as shown in Fig.~\ref{fig:decay_rates}.

The diameter and temperature dependence of $\gamma_\rad$ can be obtained from the ratio of the PL intensity and the total decay time $I/\tau$. At low temperatures, $\tau_\rad$ is approximately $\tau_{\rad 0}$, and we found it to be independent of QD diameter $D$ as shown in Fig.~\ref{fig:radiative_decay}(a). Since at $T \rightarrow 0$, $\gamma_\thm$ is negligible, the total decay rate in Eq.~(\ref{equ:gamma}) can be simplified into:
\begin{equation}
\gamma(D, T \rightarrow 0) =  \gamma_{\rad 0} + \gamma_\tnl = \gamma_{\rad 0} 
+ \frac{c_1}{D} \rme^{-\frac{2\sqrt{2m}}{\hbar}\int \limits_0^{D/2} \sqrt{\phi(r)} \rmd r}.
\label{equ:gamma_T0}
\end{equation}
From the $\tau(D, T \rightarrow 0)$ data in Fig.~\ref{fig:decay_rates}(a) we obtained $\gamma_{\rad 0}$ and $c_1$.
$\Delta$ is obtained from the $\tau_{\rad}(T)$ data in Fig.~\ref{fig:radiative_decay}(b), which is generated from Figs.~\ref{fig:decay_rates}(b) and (c) using Eq.~(\ref{equ:gammar_gamma_I}). 
Finally, after obtaining all the parameters related to $\gamma_\rad$ and $\gamma_\tnl$, we obtain the $\gamma_\thm$ parameters, $c_2$ and $c_3$, by fitting the $\tau(T)$ data in Fig.~\ref{fig:decay_rates}(c) using Eq.~(\ref{equ:gamma}).

In the rest of this section, we will provide further explanations for the expression of each decay channel in Eq.~(\ref{equ:gamma}) and the procedures for extracting the five parameters from Figs.~\ref{fig:decay_rates} and \ref{fig:radiative_decay}.

\subsection{Radiative decay rate}
\label{ssec:gamma_r}

\begin{figure}
\includegraphics[width=0.9\textwidth]{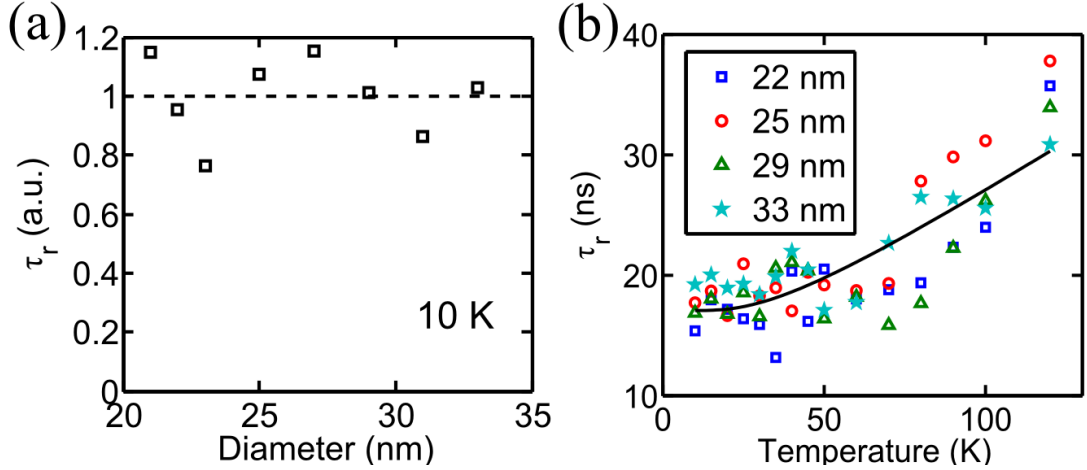}
	\caption{
	(a) The relative $\tau_\rad(D)$ extracted from the $I(D)$ and $\tau(D)$ data in Fig.~\ref{fig:decay_rates}(a) using Eq.~(\ref{equ:gammar_gamma_I}). $\tau_\rad(D)$ is normalized to the average value (dashed line).
	(b) The $\tau_\rad(T)$ relation of dense arrays with $D = 22$~nm (blue square), 25~nm (red circle), 29~nm (green triangle), 33~nm (cyan star) nm. Each $\tau_\rad$ data point is the ratio of the corresponding $\tau$ and $I$ values in Figs.~\ref{fig:decay_rates}(c) and (b), respectively, whose absolute values were originally meaningless. All four sets of data are then fitted together using Eq.~(\ref{equ:gamma_r}) with $\gamma_{\rad 0}$ and $\Delta$ being the fitting parameters, resulting in $\Delta = 8.6$~meV and an meaningless $\gamma_{\rad 0}$ value. All data and the fitted curve (solid line) are then multiplied by a common factor so that $\gamma_{\rad 0} = 59$~MHz, as obtained in Fig.~\ref{fig:decay_rates}(a), and the absolute values of $\tau_\rad$ become meaningful.}
	\label{fig:radiative_decay}
\end{figure}

First, we show that the radiative decay rate $\gamma_\rad$ of our QDs is mostly independent of the QD diameter $D$ at 21~nm $<D<33$~nm. 
$\gamma_\rad$ can be calculated from the total decay rate $\gamma$ and the PL intensity per unit InGaN area $I$ using the following equation:
\begin{equation}
\gamma_\rad \propto \gamma I.
\label{equ:gammar_gamma_I}
\end{equation}
This is because, on one hand, $\gamma_\rad$ of an exciton is related to the total decay rate $\gamma$ and QE $\eta$ via $\eta = \gamma_\rad/\gamma = \gamma_\rad/(\gamma_\rad + \gamma_\mathrm{nr})$, in which $\gamma_\mathrm{nr}$ is the nonradiative decay rate; on the other hand, $\eta$ is proportional to $I$, for QDs of diameter 10~nm $<D<40$~nm excited by the same laser intensity \cite{STRAIN}. 
Applying Eq.~(\ref{equ:gammar_gamma_I}) to the $\tau(D)$ ($\tau = 1/\gamma$) and $I(D)$ data in Fig.~\ref{fig:decay_rates}(a), we obtain the relative radiative decay time $\tau_\rad$ ($\propto \tau/I$) for QD arrays of various $D$'s at 10~K, as shown in Fig.~\ref{fig:radiative_decay}(a). This figure suggests that for QDs of 21~nm $<D<33$~nm, $\tau_\rad$ is almost constant for different $D$'s with $<24\%$ fluctuations. Henceforth, we use $\gamma_{\rad 0}$ to denote $\gamma_\rad$ of all QDs of 21~nm $<D<33$~nm at 10~K. 
The value of $\gamma_{\rad 0}$ will be extracted later in Sec.~\ref{ssec:gamma_tnl}, together with $\gamma_{\mathrm{tnl}}$, from the $\tau(D)$ data in Fig.~\ref{fig:decay_rates}(a).

The above $D$-independence of $
\gamma_\rad$ at 21~nm $<D<33$~nm is not inconsistent with our previously reported drastic increase of $\gamma_\rad$ as $D$ decreased from $2~\mu$m to $40$~nm \cite{STRAIN}. In \cite{STRAIN}, the increase of $\gamma_\rad$ was due to the strain relaxation at the nanodisk sidewall \cite{Kawakami2010}, which led to the reduction in the overall polarization fields in the nanodisk and, consequently, the improvement in the exciton oscillator strength \cite{Kawakami2006, Holmes2009}. In this work, the strain is already greatly relaxed in QD of $D < 40$~nm, so that further reduction in $D$ does not significantly improve the oscillator strength any more. 

Second, we show that the radiative decay rate $\gamma_\rad$ decreases slightly with increasing temperature due to the scattering and thermalization processes \cite{Feldmann1987}, as shown by the increase of $\tau_\rad$ in Fig.~\ref{fig:radiative_decay}(b). In an ideal 2D system, free of impurity-, phonon- and interface-scatterings, an exciton only radiatively recombines if its in-plane momentum $k = 0$, or equivalently, kinetic energy $E_k=0$. In the presence of various scatterings, characterized by a homogeneously broadened linewidth $\Delta$, an exciton with $k = 0$ may be scattered away from $k=0$, leading to a reduced recombination probability $\propto 1/\Delta$, whereas an exciton with $E_k < \Delta$ can be scattered into $k = 0$. According to Boltzmann distribution, the probability for an exciton to have $E_k < \Delta$ is $(1-\rme^{-\Delta/k_BT})$. Therefore, the $T$-dependence of $\gamma_\rad$ follows $\gamma_\rad \propto (1-\rme^{-\Delta/k_BT})/\Delta$ \cite{Feldmann1987}. The linewidth $\Delta$ depends on $T$ as $\Delta = \alpha + \beta T$ in the temperature range studied here, with $\alpha$ due to the impurity- and interface-scatterings, and $\beta T$ due to the acoustic-phonon-scattering.
The $\tau_\rad(T)$ data in Fig.~\ref{fig:radiative_decay}(b) shows a non-linear $T$-dependence of $\tau_\rad$, suggesting that $\Delta$ is NOT much less than $k_B T$. On the other hand, $\beta T \ll k_B T$, since $\beta \sim 1.7$~$\mu$eV/K for InGaN QDs \cite{Seguin2004}. Therefore, we have $\beta T \ll \alpha$, i.e. impurity- and interface-scatterings are the dominant scattering mechanisms in our QDs. Henceforth, we treat $\Delta$ as $T$-independent, which leads to a simplified $\gamma_\rad$ expression:
\begin{equation}
\gamma_\rad = \gamma_{\rad 0}(1-\rme^{-\Delta/k_BT}).
\label{equ:gamma_r}
\end{equation}
Using this equation to fit the $\tau_\rad(T)$ data in Fig.~\ref{fig:radiative_decay}(b), we obtained $\Delta = 8.6$~meV.

The above discussion on the $T$-dependence of $\gamma_\rad$ assumed that the PL of our QDs is dominated by the free-exciton emission as opposed to localized- or bound-exciton emissions. This is supported by the sharp cutoff of the $I(D)$ data at $D \approx 20$~nm, as shown in Fig.~\ref{fig:decay_rates}(a). In our sample, $>90\%$ of QDs with $D>25$~nm are optically active, while hardly any QDs with $D<19$~nm are. Such high sensitivity of $I$ to $D$ suggests that the surface nonradiative channels on the nanodisk sidewall dictate the exciton decay process, and that excitons are not confined by a-few-nanometer-scale localization centers or impurities, but by the entire nanodisk. 

\subsection{Tunnelling decay rate}
\label{ssec:gamma_tnl}

At temperature $T\rightarrow 0$, the dominant surface nonradiative recombination is due to the tunnelling of excitons through the potential barrier to the sidewall surface. The rate of the tunnelling decay is determined by the coupling of the exciton and the sidewall surface states. A rigorous calculation of this recombination rate requires the full knowledge of the wave-functions of the exciton and surface states as well as the coupling Hamiltonian, which are difficult to obtain. 
Alternatively we evaluate such a pure quantum-mechanical nonradiative decay by a phenomenological semi-classical model. We calculate the probability for an exciton with zero kinetic energy to tunnel to the sidewall through the potential barrier $\phi(r)$ using the Wentzel-Kramers-Brillouin (WKB) approximation along the nanodisk radius and write the tunnelling decay rate $\gamma_{\mathrm{tnl}}$ as:
\begin{equation}
\gamma_{\mathrm{tnl}} = \frac{c_1}{D} \rme^{-\frac{2\sqrt{2m}}{\hbar}\int \limits_0^{D/2} \sqrt{\phi(r)} \rmd r}.
\label{equ:gamma_tnl}
\end{equation}
Here, the scaling constant $c_1$ is proportional to the probability of an exciton at the surface to recombine with surface states, for which we neglected its temperature dependence. The $1/D$ factor is the sidewall surface-to-volume-ratio taking into account that the tunnelling happens along the entire nanodisk circumference $\pi D$ and that the exciton is distributed over the entire area $\pi D^2/4$. The potential barrier $\phi(r)$ was obtained in Sec.~\ref{sec:potential}. 

Neglecting $\gamma_\thm$ at low $T$, we can obtain the parameter $c_1$ together with $\gamma_{\rad 0}$ from the $\tau(D, T \rightarrow 0)$ data in Fig.~\ref{fig:decay_rates}(a), using Eq.~(\ref{equ:gamma_T0}), as mentioned earlier. The fitting yields $c_1 = 15$~m/s and $\gamma_{\rad 0} = 59$~MHz.

As $T$ increases, $\gamma_{\mathrm{tnl}}$ increases due to the occupation of states with higher kinetic energies. Excitons in higher kinetic energy states see effectively lower potential barriers and thus tunnel faster. However, to simplify the discussion, we only retain the $T$-independent part of the tunnelling decay rate in $\gamma_{\mathrm{tnl}}$ and include the $T$-dependent part into the thermal decay rate $\gamma_{\mathrm{thm}}$ by lowering the effective barrier height for thermal decay, as discussed next.

\subsection{Thermal decay rate}
\label{ssec:gamma_thm}

At a given temperature $T$, an exciton has a probability $\rme^{-\phi_B/k_BT}$ to gain a kinetic energy greater than the potential barrier height $\phi_B$ defined in Sec.~\ref{sec:potential}. Such an exciton can overcome the potential barrier without tunnelling and travel with thermal velocity $v_B$ towards the sidewall surface to recombine nonradiatively. Therefore, the thermal decay rate $\gamma_{\mathrm{thm}}$ can be written as:
\begin{equation}
\gamma_{\mathrm{thm}} = \frac{c_2 v_B}{D} \rme^{-c_3\phi_B/k_BT},
\label{equ:gamma_thm}
\end{equation}
in which, $c_2$ is a proportionality factor, and $1/D$ is again the sidewall surface-to-volume-ratio explained in Eq.~(\ref{equ:gamma_tnl}), $v_B = \sqrt{2 (c_3 \phi_B + k_BT)/m}$ is the average thermal velocity derived using the Boltzmann distribution. Note that we lowered the effective barrier height by multiplying a factor $c_3$ ($<1$) with the $\phi_B$ in Eq.~(\ref{equ:gamma_thm}) to include the $T$-dependent part of the tunnelling decay rate, as discussed in Sec.~\ref{ssec:gamma_tnl}. 
Combining Eqs.~(\ref{equ:gamma_r}), (\ref{equ:gamma_tnl}) and (\ref{equ:gamma_thm}), we obtain the total decay rate $\gamma$ as Eq.~(\ref{equ:gamma}).


Equation~(\ref{equ:gamma}) fits the $\tau(D, T)$ data very well as shown in Fig.~\ref{fig:decay_rates}(c). The fitting gives $c_2 = 2 \times 10^{-3}$ and $c_3 = 0.33$, whereas all other parameters in Eq.~(\ref{equ:gamma}) have been obtained in previous sections.

\subsection{Discussions}
\label{ssec:discussion}

At this point, we have established a quantitative model to describe the single-exciton potential profile (Eq.~(\ref{equ:phi}), Sec.~\ref{sec:potential}) and the single-exciton decay processes (Eq.~(\ref{equ:gamma}), Sec.~\ref{sec:decay_rates}) in an InGaN/GaN QD. All parameters needed in the model were obtained from the measured optical properties of QD arrays, each containing QDs with very similar structures. These parameters are summarized in Table~\ref{tab:parameters}.

\begin{table} 
\caption{Summary of parameters of the exciton dynamics model extracted from the QD-array data in Secs.~\ref{sec:potential} and \ref{sec:decay_rates}. The uncertainty of each value represents the 95~\% confidence interval of the corresponding fitting.}
\centering 
\begin{tabular}{c c c c c} 
\hline
\hline 
Parameter & Value & Unit & Equation & Physical meaning \\ 
\hline 
$E_0$ 				& $2.93 \pm 0.02$ & eV & (\ref{equ:E_vs_D}) & Exciton energy in unstrained QW \\ 
$B_m$ 				& $477 \pm 26$ & meV & (\ref{equ:phi}) & Strain-induced redshift in a planar QW \\ 
$\kappa$			& $0.037 \pm 0.007$ & nm$^{-1}$ & (\ref{equ:phi}) & $1/\kappa$: strain-relaxed region width in a nanodisk \\ 
$\gamma_{\rad 0}$ 	& $59 \pm 24$ & MHz & (\ref{equ:gamma}), (\ref{equ:gamma_r}) & Radiative decay rate at $T\rightarrow 0$ \\ 
$\Delta$			& $8.6 \pm 1.4$ & meV & (\ref{equ:gamma}), (\ref{equ:gamma_r}) & Scattering-induced linewidth \\ 
$c_1$				& $15 \pm 7$ & m/s & (\ref{equ:gamma}), (\ref{equ:gamma_tnl}) & Surface nonradiative quality \\
$c_2$				& $0.0020 \pm 0.0008$ & none & (\ref{equ:gamma}), (\ref{equ:gamma_thm}) & Surface recombination probability \\
$c_3$				& $0.33 \pm 0.06$ & none & (\ref{equ:gamma}), (\ref{equ:gamma_thm}) & Potential-height factor for $\gamma_\thm$ \\
\hline 
\end{tabular} 
\label{tab:parameters} 
\end{table}

However, there is always unavoidable structural variations from QD to QD within the same array, which result in variations in the optical properties of individual QDs, including their PL energy, intensity, decay time and photon-antibunching properties. We will address these issues in the next two sections.

\section{Single dot PL properties}
\label{sec:statistics}

Compared to self-assembled InGaN QDs, our QDs have significantly improved control over all key structural parameters: the nanodisk thickness $l$, the indium mole fraction $x$ and the disk diameter $D$. Therefore, they have significantly reduced inhomogeneities in optical properties compared to self-assembled ones made of the same material.

However, inhomogeneities cannot be completely eliminated. There still exist finite fluctuations in all three parameters: $\delta l = 2$~monolayers (MLs) limited by MOCVD growth, $\delta D = 2$~nm limited by the electron-beam lithography and plasma etching processes, and $\delta x = 0.2~\%$ limited by the Poisson distribution of the number of indium atoms, as we have analyzed in \cite{Zhang2013b}.

To study the influence of the structural fluctuations to the inhomogeneities in optical properties, we measured the PL energy, intensity, decay time, spectral linewidth and photon-antibunching of 30 QDs with the same nominal diameter of 29~nm. We find that, despite large variations of each PL property, correlations exist among all properties. We can successfully explain these correlations and account for the inhomogeneities of the QDs by varying only the potential barrier height $\phi_B$ defined in Eq.~(\ref{equ:phi_B}).

\subsection{Inhomogeneities in PL properties}
\label{ssec:inhomogeneity}

\begin{figure}
\includegraphics[width=0.9\textwidth]{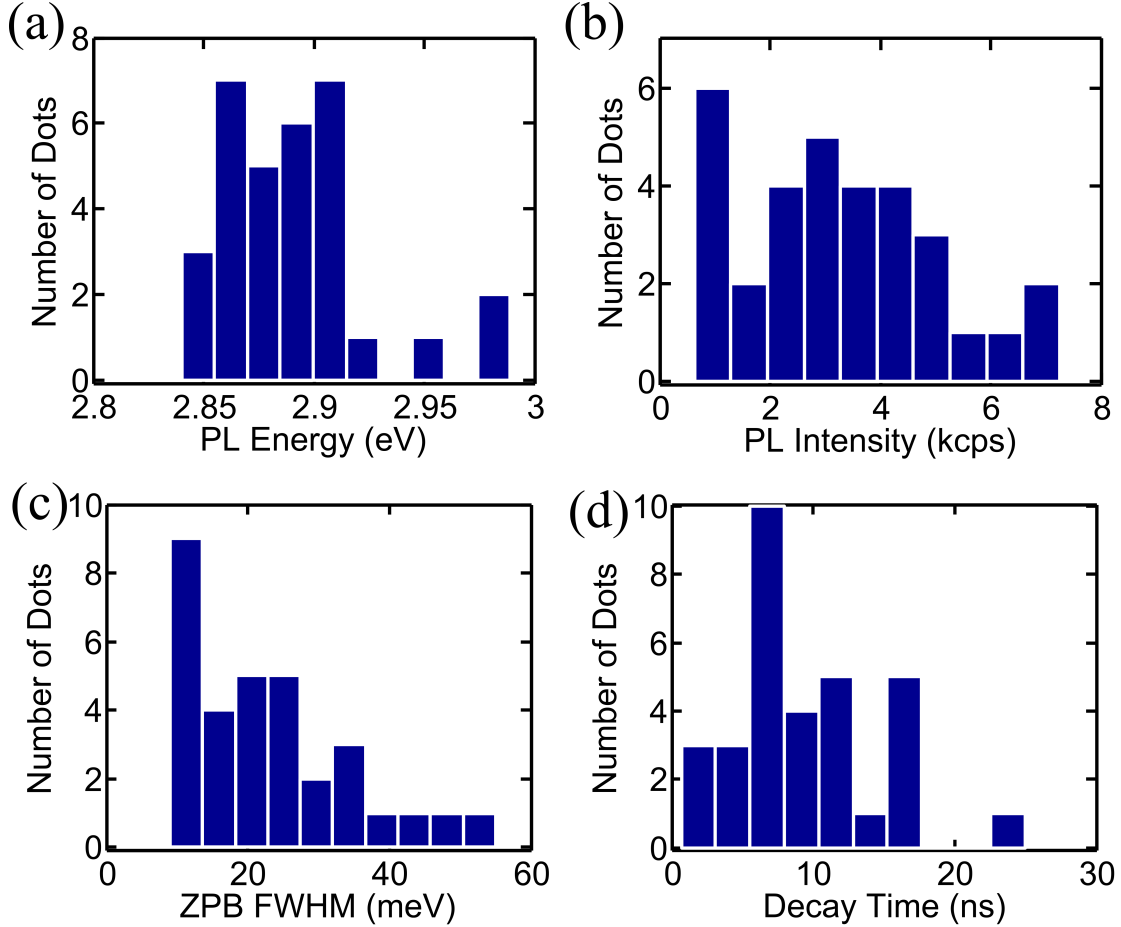}
	\caption{The statistical distributions of the PL energy $E$, the PL intensity $I$ per unit InGaN area, the FWHM $\Delta E$ of the ZPB and the PL decay time $\tau$ of 30 QDs randomly chosen from the same array of diameter $D=29$~nm. All data were taken at 10~K. The PL intensity unit kcps stands for kilo-count per second recorded by the end detector.
	}
	\label{fig:qd_statistics}
\end{figure}

To statistically characterize the inhomogeneity, we measured the PL spectra and TRPL decay traces of 30 QDs with the same nominal diameter $D = 29$~nm at 10~K. The PL spectrum of every QD consists of a dominant zero-phonon-band (ZPB) and several optical-phonon-replicas with $\sim 90$~meV interval \cite{Zhang2013b} (also see Fig.~\ref{fig:g2_vs_T}(a) for an example). However, the peak-energy $E$ of the ZPB, the integrated PL intensity $I$, the full-width-at-half-maximum (FWHM) $\Delta E$ of the ZPB, and the decay time $\tau$ vary among QDs. We plot their distributions in Fig.~\ref{fig:qd_statistics}.

As we have analyzed in \cite{Zhang2013b}, the amount of variation in $E$ matches with the estimated total amount caused by $\delta l$, $\delta x$ and $\delta D$, with $\delta l$ being the dominant source.
The influence of $l$ fluctuation to $E$ can be understood using a simple capacitor model: $E = E_0 - Fl$ \cite{Zhang2013b}, in which $F$ is the strain-induced electrical field. Hence, based on Eqs.~(\ref{equ:E_vs_D}) and (\ref{equ:phi_B}), $\delta E$ can be described by the fluctuation in $\phi_B$.
According to Fig.~\ref{fig:qd_statistics}(a), $E$ varies by about $120$~meV in the same $D=29$~nm array, which allows us to estimate that $\phi_B$ varies from $\sim 0$ to 120~meV based on Eqs.~(\ref{equ:E_vs_D}) and (\ref{equ:phi_B}).
In the following we show that indeed, for a given $D$, the influence of structural parameter fluctuation to the optical properties can be modeled by varying only the potential barrier height $\phi_B$.

\subsection{Correlations among PL properties}
\label{ssec:correlation}

\begin{figure}
\includegraphics[width=0.9\textwidth]{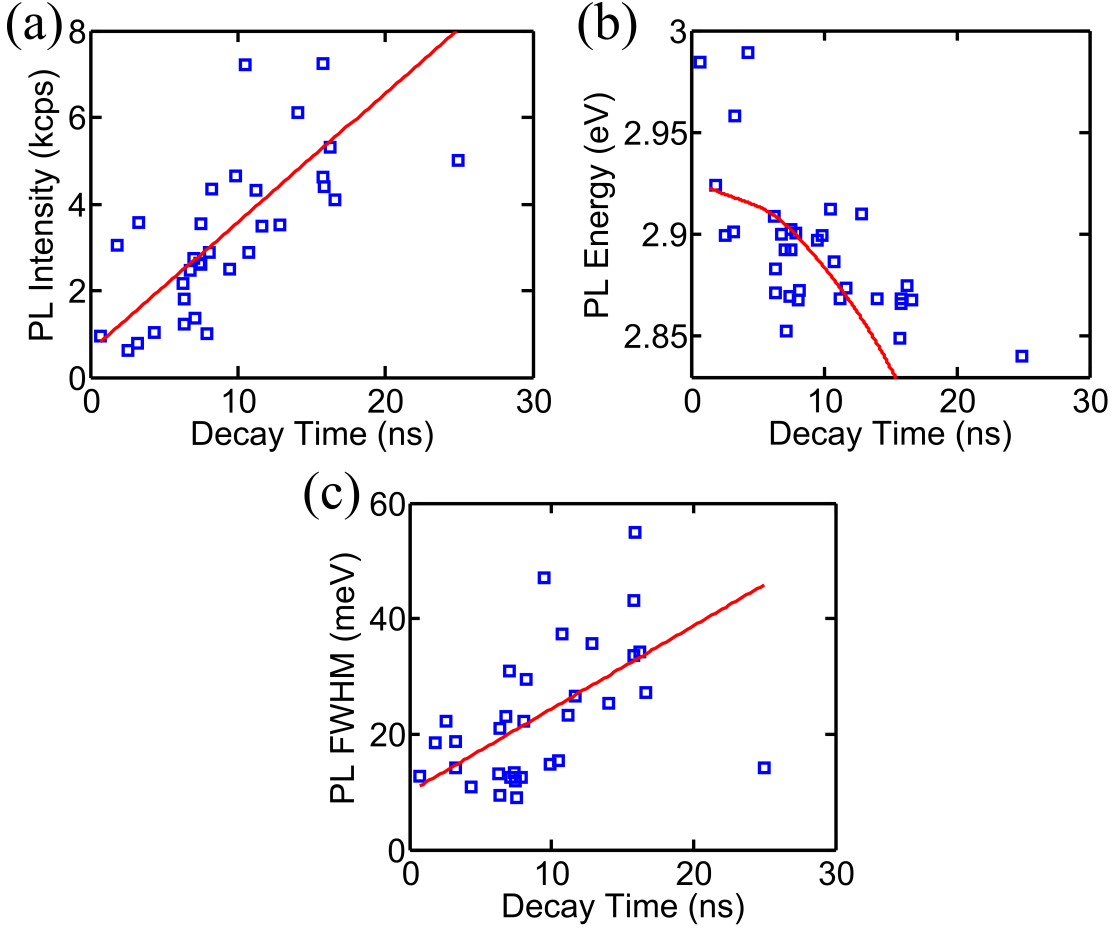}
	\caption{
	Statistical correlations among PL energy $E$, intensity $I$, decay time $\tau$ and the FWHM $\Delta E$ of the ZPB taken from 30 QDs randomly chosen from the same array of dot diameter $D=29$~nm. All data are taken at 10~K. (a) The $I$-$\tau$ correlation. Each data point represents the data from one QD. The solid line is a linear fit showing an excellent linearity between $I$ and $\tau$. 
	(b) The $E$-$\tau$ correlation. The solid line is the theoretical curve obtained by sweeping the $\phi_B$ value as described in Sec.~\ref{ssec:correlation}.
	(c) The $\Delta E$-$\tau$ correlation. The solid line is a fitting using Eq.~(\ref{equ:FWHM_vs_tau}).
	}
	\label{fig:qd_correlation_pl}
\end{figure}

Despite the seemingly random fluctuations in each PL property, we find that all measured PL properties are strongly correlated.
Figure~\ref{fig:qd_correlation_pl} shows the correlations among $E$, $I$, $\Delta E$ and $\tau$. These correlations demonstrate that all QDs in the same array share the same radiative decay time $\tau_\rad$ and that the observed PL inhomogeneities can all be modeled by the variation of $\phi_B$, as shown below.

Firstly, $\tau$ and $I$ of individual QDs are linear correlated as shown in Fig.~\ref{fig:qd_correlation_pl}(a). This suggests that $\tau_\rad$ ($\propto \tau/I$) is insensitive to the mechanism that leads to the PL inhomogeneity among QDs of the same $D$.
We have also found earlier in Sec.~\ref{ssec:gamma_r} that $\tau_\rad$ is also insensitive to $D$. Henceforth, we treat $\tau_{\rad 0}$ as a constant for all QDs of 22~nm $<D<33$~nm in our sample, whose value has already been obtained in Sec.~\ref{ssec:gamma_r} as $\tau_{\rad 0} = 17$~ns.

Secondly, all PL properties $E$, $\Delta E$ and $\tau$ are correlated as shown in Figs.~\ref{fig:qd_correlation_pl}(b) and (c). As discussed in Sec.~\ref{ssec:inhomogeneity}, the PL energy $E$ fluctuation is mainly due to the thickness fluctuation $\delta l$, or equivalently $\delta \phi_B$. On the other hand, $\phi_B$ also determines $\tau$ through Eq.~(\ref{equ:gamma_T0}). Therefore, we can plot the theoretical correlation curve (Fig.~\ref{fig:qd_correlation_pl}(b) solid line) between $E$ and $\tau$ by sweeping the $\phi_B$ value from 4 to 120~meV. For each $\phi_B$ value we obtain $E$ using Eq.~(\ref{equ:E_vs_D}) and $\tau$ using Eq.~(\ref{equ:gamma_T0}). The understanding of the correlation is straightforward: In a QD with low $E$, the potential barrier height $\phi_B$ is large. Therefore, the exciton decay is dominated by the radiative decay with a decay time close to $\tau_{\rad 0}=17$~ns, whereas the tunnelling and thermal decay are suppressed. In a QD with high $E$, the $\phi_B$ is low. Therefore, the exciton undergoes tunnelling and thermal nonradiative decay in addition to the radiative decay, leading to a short decay time.


The correlation between $\Delta E$ and $\tau$ in Fig.~\ref{fig:qd_correlation_pl}(c) can be understood as follows. The ZPB consists the contributions of multiple unresolved spectral lines, as manifested by the non-zero $g^{(2)}_0$ values from all our QDs (Fig.~\ref{fig:qd_correlation_g2}(a)). Therefore, $\Delta E$ is determined by the linewidth of each spectral line as well as the separation between lines. The impurity-scattering induces a broadening of $\Delta = 8.6$~meV in each spectral line as discussed in Sec.~\ref{ssec:gamma_r}. The rest of the $\Delta E$ is due to the spectral diffusion that leads to further broadening in each spectral line and the exction-exciton interaction that leads to larger separation between lines.
The spectral diffusion is caused by the interaction of the exciton's permanent dipole moment and the randomly appearing charges in the vicinity of the QD \cite{Ostapenko2010}. The permanent dipole moment is mainly due to the electron-hole separation by the strain-induced electric field. Hence, QDs with greater thickness $l$, or equivalently $\phi_B$, have stronger spectral diffusion.
Thicker $l$ (greater $\phi_B$) also leads to greater separations between different QD multi-excitonic states due to stronger repulsive exciton-exciton Coulomb interaction, as will be discussed in Sec.~\ref{ssec:bxx}. Meanwhile, greater $\phi_B$ also leads to longer $\tau$ as discussed earlier. Hence the positive correlation between $\Delta E$ and $\tau$ in Fig.~\ref{fig:qd_correlation_pl}(c) is explained. 
For simplicity, we approximate the positive correlation between $\Delta E$ and $\tau$ as a linear relation:
\begin{equation}
\Delta E = a + b \tau.
\label{equ:FWHM_vs_tau}
\end{equation}
We obtain from Fig.~\ref{fig:qd_correlation_pl}(c) that $a = 10 \pm 3.5$~meV and $b = 1.4 \pm 0.4$~meV/ns. 

\section{Biexciton dynamics and single-QD $g^{(2)}$ properties}
\label{sec:g2}

In previous sections we have focused on exciton dynamics in our QDs, which dominates the spectral properties at low excitation densities. Another important aspect of QD carrier dynamics is the multi-exciton interaction. However, studying the dynamics of multi-excitonic states, such as biexcitons, triexcitons, etc., typically requires the isolation of individual multi-excitonic spectral lines. This is challenging in QDs where the spectral lines are often not resolvable due to linewidth broadening caused by various carrier scattering mechanisms, spectral diffusion \cite{Nair2011a, Park2011a} and thermal broadening.

Instead, we investigate multi-exciton dynamics by measuring the degree of antibunching of the QD luminescence, characterized by the $g^{(2)}(t)$ function. $g^{(2)}$ reflects the biexciton-to-exciton QE ratio when the QD is weakly populated \cite{Nair2011a}, corresponding to the ratio of their total decay times as will be explained in Sec.~\ref{ssec:g2_and_qe}. Together with the exciton decay times obtained in Secs.~\ref{sec:potential}, \ref{sec:decay_rates} and \ref{sec:statistics}, this reveals the information of the biexciton decay times.

We will first show how to include biexciton dynamics in our model in Sec.~\ref{ssec:bxx} and evaluate the antibunching, or $g^{(2)}_0$, consistently in Sec.~\ref{ssec:g2_and_qe}. Then we will use this model to explain the peculiar $g^{(2)}_0$-$\tau$ correlation of multiple QDs with the same diameter at 10~K in Sec.~\ref{ssec:g2_vs_tau} and $T$-dependence of $g^{(2)}_0$ measured from one QD at temperatures from 10~K to 90~K in Sec.~\ref{ssec:g2_vs_T}. 

\subsection{Biexciton decay rates}
\label{ssec:bxx}

The state of a QD is described by the number of excitons it contains. A QD with $N$ excitons is said to be in the $N^{\mathrm{th}}$-excitonic state, denoted as $\ket{N}$. The decay of the QD follows a cascade process in which the number of excitons reduces one by one until the QD is in the ground state $\ket{0}$, i.e. the decay follows $\ket{N}\rightarrow \ket{N-1} \rightarrow \ket{N-2} \rightarrow ... \rightarrow \ket{0}$. Due to the exciton-exciton interaction and electronic state filling, the amount of energy released during $\ket{N}\rightarrow \ket{N-1}$ is usually different from that released in $\ket{N-1}\rightarrow \ket{N-2}$. If the energy released in every cascade step is in the form of a photon, the QD will exhibit a luminescence spectrum containing multiple discrete spectral lines, each corresponding to one of the cascade steps. In this work, we denote the exciton $\ket{1}$ and biexciton $\ket{2}$ states as $\ket{\mathrm{X}}$ and $\ket{\mathrm{XX}}$, respectively. The spectral lines corresponding to $\ket{\mathrm{XX}} \rightarrow \ket{\mathrm{X}}$ and $\ket{\mathrm{X}} \rightarrow \ket{0}$ are called biexciton and exciton emission, respectively. The difference between the exciton and biexciton emission energies are often referred as the biexciton binding energy.

In our QDs \cite{STRAIN}, as well as in many other InGaN/GaN QDs \cite{Schomig2004, Martin2005, Simeonov2008, Winkelnkemper2008, Bardoux2009, Sebald2011, Amloy2012, Chen2012}, the biexciton emission typically has higher energy than the exciton emission, i.e. the biexciton has a negative binding energy $-B_\XX$. This is mainly because of the enhanced repulsive exciton-exciton Coulomb interaction due to the electron-hole separation by the strain-induced electric field. 

The biexciton binding energy $B_\XX$ is positively correlated with the ZPB linewidth $\Delta E$ and decay time $\tau$ of the QD luminescence. This is because a QD with greater $\phi_B$ (thicker $l$) has further electron-hole separation, which leads to a larger biexciton energy shift due to the stronger repulsive exciton-exciton interaction and, hence, a greater $B_{\XX}$. This, together with the stronger spectral-diffusion, leads to a larger overall $\Delta E$ in QDs whose $\ket{\X}$ and $\ket{\XX}$ lines are not resolved, as explained in Fig.~\ref{fig:qd_correlation_pl}(c) in Sec.~\ref{ssec:correlation}. For simplicity, we assume that $B_{\XX}$ is proportional to the linewidth broadening $\Delta E - \Delta$, which in turn contains joint contribution from spectral diffusion and exciton-exciton interaction:
\begin{equation}
B_{\XX} = c_\XX (\Delta E - \Delta) = c_{\XX}(a + b\tau - \Delta).
\label{equ:B_XX}
\end{equation} 
The second equation is based on Eq.~(\ref{equ:FWHM_vs_tau}). Recall that $\Delta$ is the broadening caused by the impurity scattering as discussed in Sec.~\ref{ssec:gamma_r}.

The biexciton binding energy $B_\XX$ also corresponds to the difference in the exciton and biexciton potential barrier heights. This is because the strain-induced electrical field and, therefore, the repulsive exciton-exciton Coulomb interaction, is strongest at $r = 0$ and is negligible at $r = D/2$. As a result, the biexciton potential barrier height $\phi_{B,\XX}$ can be expressed as $\phi_{B,\XX} = \phi_B - B_{\XX}$, as illustrated in Fig.~\ref{fig:model}(b). Assuming that the potential profile scales with its height, we can describe the potential profile $\phi_{\XX}(r)$ for the excitons in the $\ket{\XX}$ state by modifying Eq.~(\ref{equ:phi}) as:
\begin{equation}
\phi_{\XX}(r) = \frac{\phi_{B,\XX}}{\phi_B}\phi(r).
\label{equ:phiXX}
\end{equation}

The tunnelling and thermal nonradiative decay rates of an exciton in the $\ket{\XX}$ state, $\gamma_{\tnl,\XX}$ and $\gamma_{\thm,\XX}$, are obtained by replacing $\phi(r)$ and $\phi_B$ with $\phi_{\XX}(r)$ and $\phi_{B,\XX}$ in the decay rate equations~(\ref{equ:gamma_tnl}) and (\ref{equ:gamma_thm}). 

The radiative decay rate of an exciton in the $\ket{\XX}$ state $\gamma_{\rad,\XX}$ is assumed to be the same as that of an exciton in the $\ket{\X}$ state $\gamma_{\rad, \X}$. This is because the radiative decay rate is insensitive to the piezoelectric field and potential barrier height for 22~nm~$<D<33$~nm, as shown in Fig.~\ref{fig:decay_rates}(a) and \ref{fig:qd_correlation_pl}(a). Note that the decay rate of the $\ket{\XX}$ state is twice as fast as the decay rate of an exciton in the $\ket{\XX}$ state.

The biexciton QE $\eta_\XX$ is determined by its radiative decay rate $\gamma_{\rad,\XX}$ and nonradiative decay rate $ \gamma_{\mathrm{nr}, \XX}=\gamma_{\tnl,\XX} + \gamma_{\thm,\XX}$ via $\eta_\XX = \gamma_{\rad,\XX}/(\gamma_{\rad,\XX}+\gamma_{\mathrm{nr}, \XX})$. As we shall show next, the biexciton QE $\eta_\XX$ together with the exciton QE $\eta_\X$ determines the degree of antibunching of the QD emission.

\subsection{The relation between $g^{(2)}_0$ and $\eta_\XX/\eta_\X$}
\label{ssec:g2_and_qe}

The second-order correlation function $g^{(2)}(t)$ is defined as:
\begin{equation}
g^{(2)}(t) = \frac{\langle I_1(t')I_2(t' + t) \rangle}{\langle I_1(t')\rangle \langle I_2(t') \rangle}, 
\end{equation}
in which $I_1(t')$ and $I_2(t')$ are the photon flux intensities at time $t'$ detected by the two arms of the HBT interferometer shown in Fig.~\ref{fig:sample_and_optics}(c), and $\langle \rangle$ stands for the average over $t'$. For an ideal single-photon source, it is impossible for the two arms to detect photons simultaneously, therefore, we have $I_1(t')I_2(t')=0$ and $g^{(2)}(0) = 0$. 

When a single-photon source is triggered by periodic pulses, its $g^{(2)}$ function has a shape similar to Fig.~\ref{fig:g2_vs_T}(b-d). To characterize the number of photons emitted after each pulse, we define the quantity $g^{(2)}_0$ as:
\begin{equation}
g^{(2)}_0 = \frac{\int \limits_{0^{th}~peak} g^{(2)}(t) \mathrm{d}t}{\int \limits_{i^{th}~peak,~i>1}g^{(2)}(t) \mathrm{d}t}.
\label{equ:g2_def}
\end{equation}
In most experimental systems, due to the low photon extraction and detection efficiencies ($\ll 1$), this expression can be simplified as \cite{Nair2011a}: $g^{(2)}_0 = \langle n(n-1) \rangle/\langle n \rangle^2$, where $n$ is the number of photons emitted after each pulse.

Obviously, Eq.~(\ref{equ:g2_def}) is only applicable when the laser pulse period is much longer than the decay time of the QD so that the $g^{(2)}$ peaks do not overlap. In many of our QDs, however, the decay time is comparable with the 12.5~ns laser pulse period so that the $g^{(2)}$ peaks overlap considerably with each other, rendering Eq.~(\ref{equ:g2_def}) inapplicable. In this case, $g^{(2)}(t)$ data need to be compared with multi-exciton cascade decay model \cite{Zhang2013b} to obtain the $g^{(2)}_0$ value. When the QD is weakly populated, $g^{(2)}(t)$ only depends on the exciton total decay time $\tau_\X$ and the biexciton-to-exciton QE ratio $\eta_\XX/\eta_\X$ \cite{Zhang2013b}. Therefore, we can extract $\tau_\X$ and $\eta_\XX/\eta_\X$ from the measured $g^{(2)}(t)$ data, use $\tau_\X$ and $\eta_\XX/\eta_\X$ to reconstruct the $g^{(2)}(t)$ function of a much longer pulse period, and then calculate $g^{(2)}_0$ using Eq.~(\ref{equ:g2_def}).

We found that, in our QDs, the difference between the $g^{(2)}_0$ and $\eta_\XX/\eta_\X$ values extracted using the method described in the last paragraph was less than 0.05, approximately the error-bar of our $g^{(2)}_0$ fitting, even at the highest laser excitation power used in this section. This is expected when the average number of excitons in the QD is much less than one \cite{Nair2011a}. Henceforth, we will not distinguish $g^{(2)}_0$ and $\eta_\XX/\eta_\X$, i.e.
\begin{equation}
g^{(2)}_0 = \eta_\XX/\eta_\X.
\label{equ:g2}
\end{equation}


\subsection{The correlation between $g^{(2)}_0$ and $\tau$}
\label{ssec:g2_vs_tau}

\begin{figure}
\includegraphics[width=0.9\textwidth]{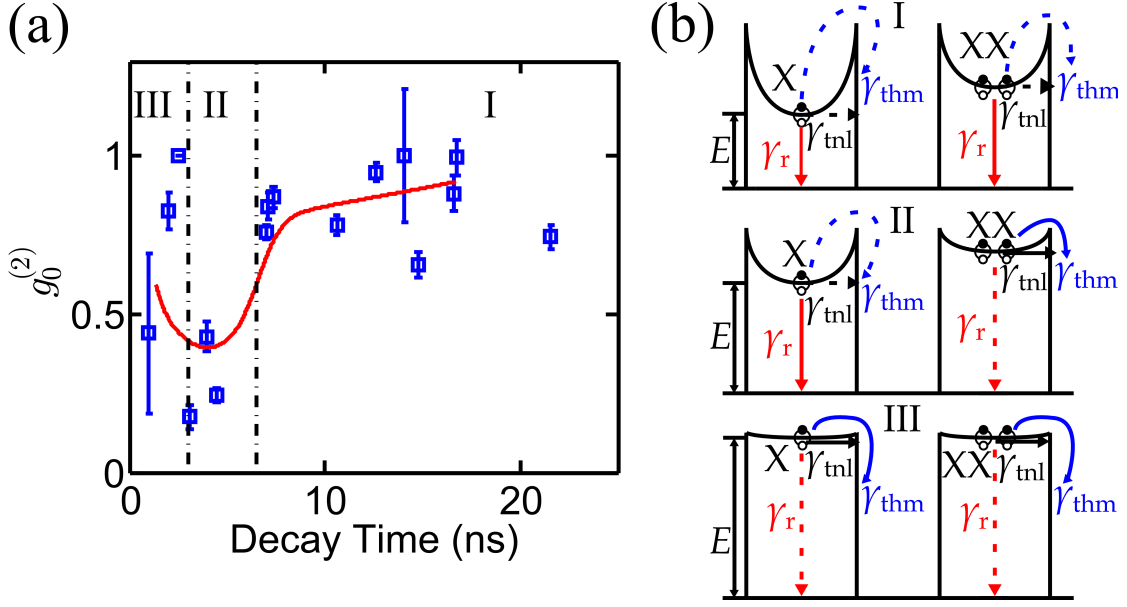}
	\caption{(a) The $g^{(2)}_0$ vs. decay time $\tau$ data of 16 randomly chosen QDs of diameter $D = 29$~nm. The solid line is obtained by sweeping the value of $\phi_B$ as explained in Sec.~\ref{ssec:g2_vs_tau}. (b) An illustration of the different physics in the three regions described in Sec.~\ref{ssec:g2_vs_tau}. The solid arrows represent dominant decay channels, whereas the dashed arrows represent less dominant channels.}
	\label{fig:qd_correlation_g2}
\end{figure}

We can apply the biexciton decay rates and the $g^{(2)}_0$ theory developed above to explain the correlation between $g^{(2)}_0$ and the total decay time $\tau$ of QD PL shown in Fig.~\ref{fig:qd_correlation_g2}(a), which is taken from 16 randomly chosen QDs of $D = 29$~nm at 10~K. 

As discussed in Sec.~\ref{sec:statistics}, the variations in PL properties among these QDs can be accounted for by the variation of a single parameter $\phi_B$ from 0 to 120~meV. For a given $\phi_B$ value, we can calculate $\gamma_{\mathrm{tnl}, \X}$ and $\gamma_{\mathrm{thm}, \X}$ for excitons using Eqs.~(\ref{equ:gamma_tnl}) and (\ref{equ:gamma_thm}) with all other parameters obtained in Secs.~\ref{sec:potential} and \ref{sec:decay_rates} as summarized in Table~\ref{tab:parameters}. This gives us $\tau_\X$ ($=\tau_{\rad, \X} + \gamma_{\mathrm{thm}, \X} + \gamma_{\mathrm{tnl}, \X}$) as well as $\eta_{\X}$ ($=\gamma_{\rad, \X}/\gamma_\X$). Due to the low excitation intensity $P$, the biexciton contributes little to the total decay time $\tau$, therefore, $\tau_\X$ can be treated as $\tau$. Each $\tau$ value corresponds to a $B_\XX$ according to Eq.~(\ref{equ:B_XX}). Knowing $B_\XX$, we can calculate the $\phi_{\XX}$, $\phi_{B,\XX}$ according to Eq.~(\ref{equ:phiXX}), from which we obtain the $\gamma_{\tnl,\XX}$, $\gamma_{\thm,\XX}$ and $\eta_{\XX}$ as described in Sec.~\ref{ssec:bxx}. Finally, $\eta_\XX/\eta_\X$ gives us $g^{(2)}_0$ (Eq.~(\ref{equ:g2})).
The only unknown parameter is the $c_\XX$ in Eq.~(\ref{equ:B_XX}) used to obtain $B_\XX$, which reflects the contribution of $B_\XX$ to the ZPB linewidth $\Delta E$. The best matching between theory and experiment is obtained when $c_\XX = 0.9$, as shown by the solid line in Fig.~\ref{fig:qd_correlation_g2}(a), suggesting that $B_\XX$ contributes significantly to $\Delta E$.

The ladle-shaped $g^{(2)}_0$-$\tau$ correlation results from the variation of the $\eta_{\XX}/\eta_{\X}$ ratio from dot to dot. The correlation curve reveals three regions I, II and III, as illustrated in Fig.~\ref{fig:qd_correlation_g2}(b), where different exciton and biexciton decay mechanisms dominate. In Region I, QDs have large $\tau$, indicating high potential barriers $\phi_B$, so that both the exciton and biexciton mainly decay radiatively and their QEs are both close to one, resulting in $g^{(2)}_0$ close to one. In Region II, QDs have medium $\tau$, suggesting relatively lower $\phi_B$. In this region, excitons still experience high enough potential barriers so that they mainly decay radiatively with high $\eta_\X$; whereas biexcitons' potential barriers are not high enough, due to the exciton-exciton Coulomb interaction, so that they mainly decay nonradiatively with low $\eta_\XX$. This leads to low $\eta_{\XX}/\eta_{\X}$ and, consequently, strong anti-bunching in $g^{(2)}$. In Region III, QDs have very short $\tau$, suggesting that the potential barriers for both excitons and biexcitons are low, so that both mainly decay nonradiatively with low $\eta_X$ and $\eta_\XX$, while their  ratio $\eta_{\XX}/\eta_{\X}$ increases.

\subsection{The temperature dependence of $g^{(2)}_0$}
\label{ssec:g2_vs_T}

\begin{figure}
\includegraphics[width=0.9\textwidth]{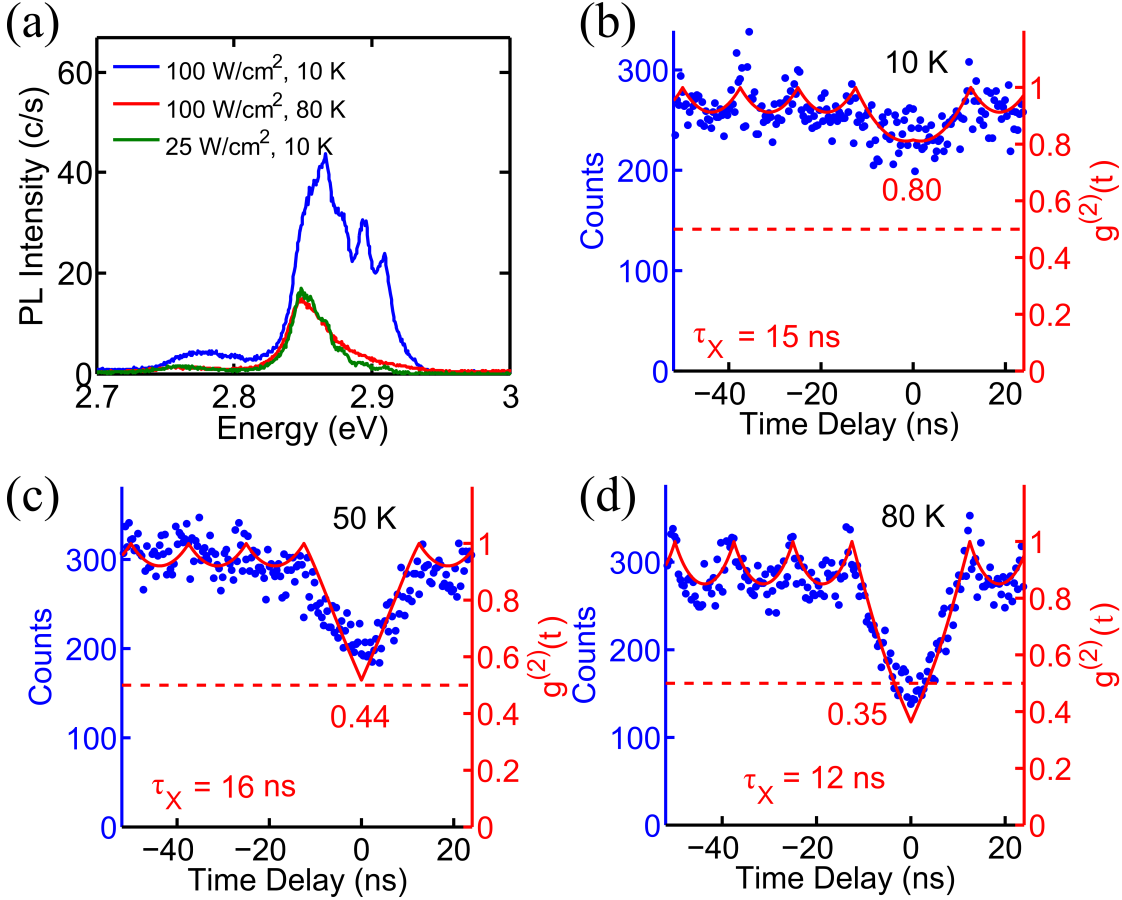}
	\caption{
	(a) The PL spectra of a QD of $D=29$~nm at $P = 100$~W/cm$^2$, $T=10$~K (blue); $P = 100$~W/cm$^2$, $T=80$~K (red); and $P = 25$~W/cm$^2$, $T=10$~K (green). (b-d) The scattered data points form the measured $g^{(2)}(t)$ of the QD at 10, 50 and 80~K, respectively, at $P = 100$~W/cm$^2$. All data are without background subtraction. The solid lines are the fitting curves obtained from the method developed in \cite{Zhang2013b} (also briefly described in Sec.~\ref{ssec:g2_and_qe}). Each fitting returns the biexciton-to-exciton QE ratio $\eta_\XX/\eta_\X$, exciton total decay time $\tau_{\X}$ and $g^{(2)}_0$. The fitted $g^{(2)}_0$ values are shown next to the central $g^{(2)}$ peaks (dips).}
	\label{fig:g2_vs_T}
\end{figure}

The difference in exciton and biexciton's potential barrier heights also leads to a peculiar temperature dependence of $g^{(2)}_0$ of single QDs. We found that, as shown in Fig.~\ref{fig:g2_vs_T}(b-d), the single-photon purity of QD emission may improve with increasing temperature.

Figure~\ref{fig:g2_vs_T}(a) shows the PL spectra of a QD of diameter $D = 29$~nm. 
At a low temperature (10~K) and excitation intensity $P=100$~W/cm$^2$, the PL spectrum is composed of multiple overlapping peaks with an overall ZPB linewidth of 50~meV. 
Keeping the same temperature but lowering $P$ weakens the higher energy peaks, suggesting that the higher energy peaks are due to multi-exciton emissions. 
Keeping the same $P$ but increasing the temperature to 80~K weakens the higher energy peaks as well, suggesting that QEs of multi-excitons reduce faster with temperature than the QE of exciton does. 

\begin{figure}
\includegraphics[width=0.9\textwidth]{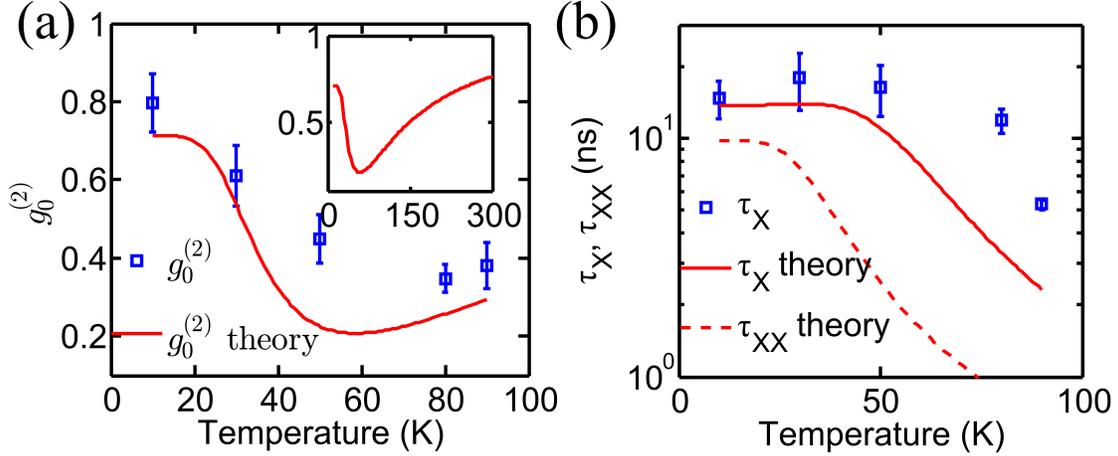}
	\caption{
	The blue squares in (a) and (b) are the $g^{(2)}_0$ and $\tau_\X$ extracted from the $g^{(2)}(t)$ data of the QD at 10, 30, 50, 80 and 90~K, three of which are shown in Fig.~\ref{fig:g2_vs_T}(b-d). The red lines are the theoretical curves based on the $\phi_B$ and $B_\XX$ values extracted from Fig.~\ref{fig:g2_vs_T}(a) as described in the text. The inset in (a) is an extension of the theoretical $g^{(2)}_0(T)$ curve to 300~K.}
	\label{fig:g2_vs_T_fitting}
\end{figure}

This mechanism is verified by the improvement of the antibunching in the $g^{(2)}(t)$ as the temperature increases from 10 to 80~K at a fixed $P$, as shown in Fig.~\ref{fig:g2_vs_T}(b-d). From each of the $g^{(2)}(t)$ data we obtain $g^{(2)}_0$ and the exciton decay time $\tau_\X$ using the method developed in \cite{Zhang2013b} and briefly summarized in Sec.~\ref{ssec:g2_and_qe}. The resulting $g^{(2)}_0(T)$ and $\tau_\X(T)$ data are shown in Figs.~\ref{fig:g2_vs_T_fitting}(a) and (b), respectively. Both sets of data can be reproduced by our model using parameters in Table~\ref{tab:parameters} as follows: 
From the 10~K PL energy $E=2.85$~eV and ZPB linewidth $\Delta E = 50$~meV (Fig.~\ref{fig:g2_vs_T}(a), $P = 100$~W/cm$^2$) we obtain $\phi_B = 80$~meV and $B_\XX = 37$~meV, according to Eqs.~(\ref{equ:E_vs_D}) and (\ref{equ:B_XX}), respectively. Following the same steps as we used in Sec.~\ref{ssec:g2_vs_tau}, $\phi_B$ gives rise to $\gamma_\X(T)$ and $\eta_\X(T) = \gamma_{\rad,\X}(T)/\gamma_\X(T)$ through Eq.~(\ref{equ:gamma}); whereas $B_\XX$ gives rise to $\phi_{B,\XX}$ based on Eq.~(\ref{equ:phiXX}), which leads to $\gamma_\XX(T)$ and $\eta_\XX(T) = \gamma_{\rad, \XX}(T)/\gamma_\XX(T)$ through Eq.~(\ref{equ:gamma}) as well. Using Eq.~(\ref{equ:g2}) and recalling the assumption that $\gamma_{\rad,\X} = \gamma_{\rad,\XX}$ in Sec.~\ref{ssec:bxx}, the ratio $\eta_\XX(T)/\eta_\X(T) = \tau_\XX(T)/\tau_\X(T)$ gives rise to the theoretical $g^{(2)}_0(T)$ curve shown as the solid line in Fig.~\ref{fig:g2_vs_T_fitting}(a). For comparison, the theoretical $\tau_\X(T)$ and $\tau_\XX(T)$ are plotted in Fig.~\ref{fig:g2_vs_T_fitting}(b). From Fig.~\ref{fig:g2_vs_T_fitting}(b) it is evident that the improvement of the $g^{(2)}_0$ as temperature increases is due to the faster dropping of $\tau_\XX$ compared to $\tau_X$, i.e. the biexciton's thermal decay rate increases faster than the exciton's. This is ultimately because, in the same QD, the biexciton has a lower potential barrier than the exciton as a result of the repulsive exciton-exciton Coulomb interaction. 

However, $g^{(2)}_0$ does not always decrease with temperature. Instead, there is an optimal temperature, at which $g^{(2)}_0$ reaches its minimal. In the inset of Fig.~\ref{fig:g2_vs_T_fitting}(a), the increase of $g^{(2)}_0$ as $T$ increases above 80~K is because the thermal decay of both exciton and biexciton dominate the total decay and increase exponentially with temperature. This quickly lowers both $\eta_\X$ and $\eta_\XX$ and raises $g^{(2)}_0$, similar to what happened in the Region I of Fig.~\ref{fig:qd_correlation_g2}. The optimal temperature increases with $\phi_B$, since greater $\phi_B$ means higher QEs for both exciton and biexciton, as well as larger difference between them.

This anomalous temperature dependence of $g^{(2)}_0$ suggests that one way to improve the operating temperature of these QDs is to increase the $\phi_B$ value by, for instance, having higher indium composition or thicker InGaN layer. 

\section{Conclusion}
\label{sec:conclusion}

We studied the carrier dynamics of site- and structure-controlled nanodisk-in-a-wire InGaN/GaN QDs. The minimized inhomogeneities in all key structural parameters--the QD diameter, thickness and indium composition--allowed a systematical mapping between optical properties and structural parameters.

Our results revealed that the sidewalls in these etched QDs played a vital role in enhancing the radiative decay rate and enabling good antibunching, while it also ultimately limited the QE. The strain relaxation at the sidewall led to greatly enhanced radiative decay rates in QDs compared to QWs \cite{STRAIN}. More importantly, it created potential barriers, which are different for excitons and biexcitons and preferentially protected excitons from surface recombination, leading to low $g^{(2)}_0$ values. This suggests that by engineering the potential barrier height, such as by varying the indium composition or nanodisk thickness, one could achieve purer single-photon emission and at higher temperatures with QDs fabricated by our method.

However, the QE was ultimately limited by the surface recombination at the sidewall, even at very low temperatures, due to tunneling of the carriers to the sidewall. Such surface dynamics has often been overlooked in dot-in-a-nanowire based optical devices, especially in low-temperature measurements.

Furthermore, the statistical correlations between various optical properties of numerous single QDs with markedly similar diameters enabled us to understand the impacts of structural parameters on the optical properties. Together with the study on QD ensembles, we established a quantitative relation between the optical properties and the structural parameters. And we showed that variations in optical properties of QDs of the same diameter could be modeled by the variation of only one phenomenological parameter, the exciton potential barrier height. 

These findings may be applicable to a wide range of strained III-N nanostructures with large surface-to-volume ratios, such as nanowires, nanospheres, and nanopillars. Hence
the carrier dynamics we analyzed and quantitatively modeled in this work may have broad impacts on improving the performance of III-N based photonic devices.

\begin{acknowledgments}
We acknowledge financial supports from the National Science Foundation (NSF) under Awards ECCS 0901477 for the work related to materials properties and device design, ECCS 1102127 for carrier dynamics and related time-resolved measurements, and DMR 1120923 (MRSEC) for work related to light-matter interactions. The work related to epitaxial growth, fabrication, and photon antibunching properties were also partially supported by the Defense Advanced Research Project Agency (DARPA) under grant N66001-10-1-4042. Part of the fabrication work was performed in the Lurie Nanofabrication Facility (LNF), which is part of the NSF NNIN network.
\end{acknowledgments}

\end{document}